\documentclass[10pt,  aps, prd, nofootinbib, twocolumn, superscriptaddress]{revtex4-2}

\emergencystretch=\maxdimen
\usepackage[mathlines]{lineno}

\usepackage{amsmath}
\usepackage{float}
\usepackage{amssymb}
\usepackage[normalem]{ulem}
\usepackage{bm}          
\usepackage{braket}
\usepackage{dcolumn}     
\usepackage{diagbox} 
\usepackage{epsfig}
\usepackage{graphicx}   
\usepackage{hhline}
\usepackage{mathrsfs}
\usepackage{newtxmath, newtxtext}
\usepackage{slashed}
\usepackage{subfigure}
\usepackage{xcolor}
\usepackage{mathtools, cuted}
\usepackage{lipsum}

\usepackage[pagebackref=false, colorlinks=true]{hyperref}
\definecolor{reddish}{rgb}{0.7,0.2,0.0}  
\definecolor{blueish}{rgb}{0.1,0.1,1}
\hypersetup{
linkcolor=reddish,      
citecolor=blueish,      
filecolor=blue,         
urlcolor=magenta}       

\newcommand\at[2]{\left.#1\right|_{#2}}

\def\apj{ApJ}                    
\def\apjl{ApJL}                  
\def\apjs{ApJS}                  
\def\ao{Appl.~Opt.}              

\def\aap{A\&A}                   
\def\cqg{Class. Quantum Grav.}   
\def\jcap{JCAP}                  
\def\mnras{MNRAS}                
 
\def\prd{Phys.~Rev.~D}           

\def\sciadv{Sci. Adv.}

\begin{document}

\title{Influence of Observer's Inclination and Spacetime Structure on Photon Ring Observables}

\author{Kiana Salehi}\email{k2salehi@uwaterloo.ca}
\affiliation{Perimeter Institute for Theoretical Physics, 31 Caroline Street North, Waterloo, ON, N2L 2Y5, Canada}
\affiliation{Department of Physics and Astronomy, University of Waterloo,200 University Avenue West, Waterloo, ON, N2L 3G1, Canada}
\affiliation{Waterloo Centre for Astrophysics, University of Waterloo, Waterloo, ON N2L 3G1 Canada}
\author{Rahul Kumar Walia}
\affiliation{Department of Physics, University of Arizona, 1118 E 4th Street, 85721 Tucson, USA}
\author{Dominic Chang}
\affiliation{Department of Physics, Harvard University, Cambridge, MA 02138, USA}
\affiliation{Black Hole Initiative at Harvard University, 20 Garden St., Cambridge, MA 02138, USA}
\author{Prashant Kocherlakota}
\affiliation{Black Hole Initiative at Harvard University, 20 Garden St., Cambridge, MA 02138, USA}
\affiliation{Center for Astrophysics, Harvard \& Smithsonian, 60 Garden St., Cambridge, MA 02138, USA}

\begin{abstract}
Recent observations of the near-horizon regions of BHs, particularly the images captured by the Event Horizon Telescope (EHT) collaboration, have greatly advanced our understanding of gravity in extreme conditions. These images reveal a bright, ring-like structure surrounding the central dark area of supermassive BHs, created by the images of unstable photon orbits. As observational capabilities improve, future studies are expected to resolve higher-order rings, providing new opportunities to test gravity through observables such as the Lyapunov exponent, time delay, and azimuthal shift. These observables offer valuable insights into the structure of spacetime, BH properties, and the inclination of the observer.
In this study, we employ a non-perturbative and non-parametric framework to examine how these observables change with deviations from the no-hair theorem and varying inclinations. We focus particularly on polar observers, which are highly relevant for M87$^*$. Our analysis explores how each of these observables can reveal information about the structure of spacetime and the morphology and existence of the ergosphere and event horizon. Furthermore, we illustrate this characterization for several specific alternative spacetimes, investigating how these current and potential future measurements, including those of the shadow size, can provide direct insights into the spin parameter values for each of these spacetimes.

\end{abstract}

\maketitle

\section{Introduction}
\label{Sec-1}

Recent advancements in horizon-scale observations of black holes (BHs) by the Event Horizon Telescope (EHT) collaboration have opened a new frontier for probing gravity in strong-field regimes \cite{EHTC+2019a, EHTC+2022a, EventHorizonTelescope:2024dhe}. The EHT has captured a bright, ring-like structure, known as the emission ring, surrounding a central brightness depression in the images of the supermassive compact objects M87$^*$ and Sgr A$^*$, located at the centers of M87 and our Milky Way galaxy, respectively.

The central brightness depression is best explained by a spacetime feature known as the photon shell, a region outside the BH%
\footnote{We note that certain horizon-less ultra-compact objects can also possess photon shells and cause central brightness depression. In the absence of spin, an extended photon shell reduces to a degenerate photon sphere.} %
in which photons are allowed to move on bound, spherical orbits of constant radii. On the observer's screen, the gravitationally-lensed image of the photon shell is called the shadow boundary curve or critical curve, and is characterized purely by the spacetime-geometry and the inclination of the observer \cite{Bardeen1973, Perlick:2021aok, Cunha:2018acu, Lupsasca:2024wkp}.
Photons that appear inside the critical curve have shorter path lengths in the bulk compared to those that appear outside \cite{Jaroszynski+1997, Falcke+2000}. 
Since photons with shorter path lengths pass through a relatively smaller amount of the hot accreting plasma on the way to the observer, a concomitant central brightness depression in the observed image is predicted. Indeed, the diameters of the observed emission rings surrounding the dark regions in the EHT images have been used to infer the sizes of the BH shadows of M87$^*$ and Sgr A$^*$ \cite{EHTC+2019f, EHTC+2022f}. These shadow sizes, in turn, have successfully been used to ``measure'' various aspects of their spacetimes \cite{Psaltis+2020, Kocherlakota+2021, EHTC+2022f,KumarWalia:2022aop,Vagnozzi:2022moj,Kumar:2020yem}.

The bound spherical photon orbits in the photon shell of a typical BH are unstable. 
Consequently, photons that are slightly displaced from these orbits follow vastly different trajectories: Nearly-bound rays with small radial deviations circle the BH a finite number of times before eventually escaping its gravitational pull and reaching a distant observer.
Such lensing causes a single point source in the bulk of space to cast a countably-infinite number of images on the observer's screen which are often indexed with $n$.
The direct or zeroth-order image ($n=0$) is formed by the photons that reach the observer without looping around the BH, and all higher-order ($n\geq 1$) images are identified as relativistic images, formed by photons which keenly feel the effect of the BH \cite{Luminet:1979nyg}. 
When the source of emission is extended, its relativistic images form a sequence of ring-like images, called ``photon subrings". 
Since only nearly-bound photons undergo such intense light bending, all subrings appear within a narrow band straddling the critical curve, called the photon ring \cite{Johnson:2019ljv}. 

Photon subrings are characterized by universal phenomena \cite{Gralla:2019drh, Johnson:2019ljv} which demagnify, rotate, and time-delay each successive subimage relative to the previous. 
These effects are captured by the purely geometrical, critical lensing parameters which are determined by the BH spacetime and the observer's inclination angle.
In recent years, photon subrings have been extensively studied in the context of Schwarzschild and Kerr BHs, focusing on their image morphology \cite{Gralla:2020yvo, Gralla:2020nwp,Vincent+2022,Kocherlakota+2024a,Kocherlakota+2024b}, interferometric signatures \cite{Gralla:2019drh,Gralla:2019drh,Johnson:2019ljv,Paugnat+2022,Cardenas-Avendano:2023dzo}, intensity auto-correlation \cite{Hadar+2021}, holographic signatures \cite{Hadar+2022}, and polarization effects \cite{Himwich:2020msm}. 
These studies suggest that photon subrings carry important information of the near-horizon geometry of BHs. 
While current EHT observations have not yet directly detected a photon ring in the images of Sgr A$^*$ and M87$^*$, future observational advancements with the Black Hole Explorer (BHEX), with improved angular resolution and sensitivity, are expected to resolve the $n=1$ subrings in their images \cite{Johnson:2024ttr, Gralla:2020nwp}.
Such photon ring detections and associated measurements of the critical parameters could offer novel insights into the strong-field features of gravity, largely independently of astrophysical uncertainties.

Most work on the photon ring in non-Kerr spacetimes has been restricted to spherical symmetry \cite{Wielgus:2021peu, Staelens:2023jgr,Kocherlakota+2024a,Kocherlakota+2024b, Salehi+2023} with few exceptions.
In Ref.~\cite{Salehi+2023}, the variation of the demagnification parameter or Lyapunov exponent with varying spacetime geometry has been explored, however, limited to observers lying along the BH spin axis. 
Here, we extend this analysis to describe all the critical parameters as functions of spacetime geometry as well as of the observer inclination. The work presented here essentially extends the framework of Ref.
\cite{Gralla:2019drh} into non-Kerr spacetimes, and fills an important gap for investigations into how photon subrings evolve with generic deviations from the Kerr spacetime. 
We adopt a general non-perturbative and non-parametric framework, which can be used to describe a large class of stationary and axisymmetric, integrable spacetimes.

We also focus on the appearance of the photon ring when the observer is present at small inclinations $\mathscr{i}$ relative to the BH spin axis, for its value in interpreting images of M87$^*$, for which $\mathscr{i} \approx 17^\circ$ \cite{CraigWalker:2018vam}. 
In particular, we find that measurements of the critical parameters by such observers can provide insights into the existence and the structure of an ergosphere as well as an event horizon around the central object. 
This highlights the need for careful considerations of the observer's inclination angle when interpreting observational data and extracting information about BH properties.

The content of this paper is organized as follows: Section \ref{Sec-2} explores spherical photon orbits within the photon shell and introduces the critical parameters governing successive photon subrings on the image plane, all within a general, non-perturbative, and non-parametric framework. 
Section III addresses the scenario of observers located along the BH's axis of rotation, which is particularly valuable for the case of M87$^*$. 
Finally, Section V provides conclusions drawn from the study.

\section{Photon Ring Critical Parameters in Stationary and Axisymmetric, Integrable Spacetimes}
\label{Sec-2}

In this section, we introduce the lensing critical parameters that control various characteristic features of photon subrings in a broad class of stationary and axisymmetric spacetimes. 
In general relativity and any metric theory of modified gravity, a general stationary, axisymmetric, and asymptotically flat vacuum BH metric can be expressed in terms of only four independent functions \cite{Wald:1984rg}. 
Several parametric spacetimes, incorporating one or more independent functions that introduce deviations from the Kerr metric, have been proposed in the literature \cite{Papadopoulos:2018nvd,Konoplya:2018arm,Carson:2020dez,Azreg-Ainou2014,Ghasemi-Nodehi:2016wao, Konoplya:2016jvv, Cardoso:2015xtj, Cardoso:2014rha, Johannsen:2011dh,Glampedakis:2017dvb}. 
Here, we focus on one such four-dimensional and geodesically-integrable spacetime metric, \textit{viz.} the Johannsen-Psaltis (JP) metric \cite{Johannsen:2013szh}. 

The JP metric is described by four free functions that parameterize potential deviations from the Kerr metric. In Boyer-Lindquist (BL; \cite{Boyer+1967}) coordinates, $x^\alpha = (t, r, \vartheta, \varphi)$, the JP metric, up to a redefinition of the free functions, takes the form \cite{Johannsen:2013szh,Salehi+2023}
\begin{align} \label{eq:JP_Metric}
\mathrm{d}s^2 =&\ 
- \frac{\Sigma}{\Pi^2}(N^2 - F^2 a^2\sin^2{\vartheta})\mathrm{d}t^2 
\nonumber \\
&\ - 2\frac{\Sigma}{\Pi^2}(rF-N^2)a\sin^2{\vartheta}~\mathrm{d}t\mathrm{d}\varphi 
\nonumber \\
&\ + \frac{\Sigma}{\Pi^2}(r^2 - N^2 a^2\sin^2{\vartheta})\sin^2{\vartheta}~\mathrm{d}\varphi^2 \nonumber \\
&\ + \frac{\Sigma B^2}{r^2 N^2}\mathrm{d}r^2 + \Sigma\mathrm{d}\vartheta^2\,,
\end{align}
with 
\begin{align}
&\Sigma(r, \vartheta) = r^2 + f(r) + a^2\cos^2{\vartheta},\\
&\Pi(r, \vartheta) = r - F a^2\sin^2{\vartheta}.
\end{align}
Here, $N, B, F$, and $f$ are four arbitrary functions of the radial coordinate $r$ alone, depending on the BH mass $M$, the dimensionless spin parameter $a/M$, and potentially one or more deviation parameters. Note, however, that the metric function $B$ can be eliminated through a change of the radial coordinate, $\mathrm{d}\rho = B(r)\mathrm{d}r$. Therefore, in appropriately chosen coordinates (e.g., $(t, \rho, \vartheta, \varphi)$), only three free functions are needed to describe deviations from the Kerr metric, within the family of JP metrics. 

These deviation functions are typically subjected to the conditions that the metric (\ref{eq:JP_Metric}) is regular outside the horizon, does not exhibit closed causal curves, asymptotically flat, and conforms to the Newtonian limit, thereby satisfying weak-field gravitational tests. Although the JP metric is not a solution to the field equations of any specific gravity theory, it can be mapped to a wide range of known non-Kerr BH solutions in general relativity and modified theories of gravity for particular choices of deviation functions. The JP metric simplifies to the Kerr metric for specific choices of metric functions and reduces to the Rezzolla-Zhidenko metric \cite{Rezzolla+2014} in the limit $a\to 0$:
\begin{equation}\label{eq:JP_Metric-1}
\mathrm{d}s^2 = \frac{\Sigma}{r^2}\Big(-N^2 \mathrm{d}t^2 +\frac{B^2}{N^2}\mathrm{d}r^2 +r^2 (\mathrm{d}\vartheta^2+\sin^2\vartheta~ \mathrm{d}\phi^2  ) \Big),
\end{equation}
where in the absence of a conformal singularity, $\Sigma\neq 0$, null geodesics are solely determined by only two independent deviation functions $N$ and $B$.

The formation of shadows and photon rings is a manifestation of the extreme bending of light around a BH. Thus, the first step in our analysis is to find the analytical solutions for null geodesics around a spinning BH described by the JP metric. Unless otherwise specified, hereafter, ``geodesic" will refer to the null geodesics of the JP metric (\ref{eq:JP_Metric}) followed by photons.

\subsection{Null Geodesics}\label{Sec-2a}
The JP metric (\ref{eq:JP_Metric}), which retains all the symmetries of Kerr spacetime, encompasses all Petrov type-D solutions and certain Petrov type-I solutions that admit a second-rank Killing tensor and the accompanying Carter constant (eg. Kerr-Sen BH, see Ref.~\cite{Papadopoulos:2020kxu}), enabling the separability of the Hamilton-Jacobi equations in all four coordinates. Geodesic integrability greatly simplifies the analysis of obtaining the lensing critical parameters. For spacetimes lacking a Carter constant, the geodesic equations are not analytically integrable, which often leads to chaotic geodesic behavior, requiring numerical methods to solve for null geodesics and to compute the critical parameters.

The tangent $k^\mu=\dot{x}^\mu=\mathrm{d}x^\mu/d\lambda$ along an arbitrary null geodesic $x^\mu(\lambda)$, with $\lambda$ as an affine parameter along it, can be obtained as being
\begin{align} \label{eq:NG_Tangent}
\Sigma\frac{\dot{t}}{E} =&\ \frac{r}{N^2}(r - F a \xi) + a\xi - a^2\sin^2{\vartheta} =: \mathscr{T}_r(r) + a^2\cos^2{\vartheta}\,, \nonumber \\
\Sigma\frac{\dot{\varphi}}{E} =&\ \frac{aF}{N^2}(r - F a \xi) - a + \xi\csc^2\vartheta =: \Phi_r(r)+ \xi\csc^2\vartheta\,, \nonumber \\
\Sigma\frac{\dot{r}}{E} =&\ \pm_r \frac{r}{B}\left[(r - F a \xi)^2 -N^2\mathscr{I}^2\right]^{1/2} =: \pm_r\sqrt{\mathscr{R}(r)}\,, \\
\Sigma\frac{\dot{\vartheta}}{E} =&\ \pm_\vartheta \left[\mathscr{I}^2 - (\xi\csc{\vartheta} - a\sin{\vartheta})^2\right]^{1/2} =: \pm_\vartheta\sqrt{\Theta(\vartheta)}\,. \nonumber
\end{align}
We have introduced here $\pm_r$ and $\pm_\vartheta$ to represent the signs of the radial and the polar coordinate velocities of the photon respectively. In the above, $E := -k_t$ is the conserved energy along the null geodesic associated with the timelike Killing vector $\partial_t$, $\xi := k_\varphi/k_t$ is related to its conserved azimuthal angular momentum $k_\varphi$ associated with the spacelike Killing vector $\partial_\varphi$, and $\mathscr{I}$ is its (non-negative) Carter constant \cite{Carter1968, Hioki2009}.%
\footnote{We will use the convention that $\mathscr{I}$ is the positive square root of $\mathscr{I}^2$. Thus, it is understood to be non-negative throughout this work.} %
The latter is also identified as the separability constant in the Hamilton-Jacobi equation. In a non-dispersive medium, a given null geodesic is characterized by a set of conserved quantities ($\xi, \mathscr{I}$). 
Another popular choice for the Carter constant $\eta$ (see, e.g., Refs. \cite{Gralla:2019drh,Gralla:2019xty,Chandrasekhar:1985kt}), also called the Carter integral, is related to the one introduced above as,%
\footnote{In spherically-symmetric spacetimes ($a=0$), all photon orbits are zero angular momentum orbits ($\xi = 0$), in which case the two Carter constants are related simply as $\eta = \mathscr{I}^2$.}
\begin{equation} \label{eq:Carter_Consts}
\eta = \mathscr{I}^2 - (\xi- a)^2\,. 
\end{equation}
While lensing observables do not change with the choice of the Carter constant ($\mathscr{I}$ or $\eta$), the two choices offer mathematical simplifications in specific contexts. The radius of the critical curve for a polar observer, as we will see below, is equal to $\sqrt{\mathscr{I}}$. 
Equatorial orbits, for which $\dot{\vartheta} = 0$, must have $\eta=0$. 
Photons passing through the equatorial plane have an instantaneous polar momentum of $|k_{\vartheta}|=\sqrt{\eta}$, thereby requiring $\eta\geq 0$. 

The functions $\mathscr{T}_r$ and $\Phi_r$ introduced in Eq.~(\ref{eq:NG_Tangent}) contain the purely $r-$dependent pieces of the time and azimuthal coordinate velocities. Furthermore, $\mathscr{R}$ and $\Theta$ correspond to the radial and polar null geodesic effective potentials respectively, which can be seen to satisfy ``energy equations'' of the form $\dot{r}^2 = \mathscr{R}$ and $\dot{\vartheta}^2 = \Theta$. 

The previous energy equations become precise through a change of the affine parameter along the geodesic. In terms of the Mino time, $\lambda_{\rm m}$, \cite{Mino2003}, obtained via the reparametrization 
\begin{equation} \label{eq:Mino_Time}
\mathrm{d}\lambda_{\rm m} = \frac{E}{\Sigma(r(\lambda), \vartheta(\lambda))}\mathrm{d}\lambda\,,
\end{equation}
the tangent to arbitrary null geodesics, $k^\mu = \mathrm{d}x^\mu/\mathrm{d}\lambda_{\rm m}$, then becomes 
\begin{equation}
k^\mu = (\mathscr{T}_r + a^2\cos^2{\vartheta},\, \pm_r\sqrt{\mathscr{R}}, \,\pm_\vartheta\sqrt{\Theta},\, \Phi_r + \xi\csc^2{\vartheta})\,.  
\end{equation}

The four coordinate velocities now yield four decoupled first-order ordinary differential equations for $x^\mu(\lambda_{\rm m})$ \cite{Chandrasekhar:1985kt, Gralla:2019drh}. 

Note that, in the JP spacetime, the geodesic equations for the coordinates $(t, r, \phi)$ differ from those in the Kerr metric, while the $\vartheta$ equation remains unchanged -- $\Theta(\vartheta)$ is identical in both metrics due to their asymptotic flatness \cite{Papadopoulos:2018nvd, Johannsen:2013szh}. This greatly simplifies the lensing calculation in JP spacetimes. As a result, photons with given values of impact parameters ($\mathscr{I}, \xi$) in the Kerr and JP spacetimes share identical polar turning points (see Eq.~\ref{eq:thetaTurning}) but traverse different radial and azimuthal paths in between. Therefore, the Mino time required to complete one polar orbit (in Eq.~\ref{eq:MinoTime}) is identical in both Kerr and JP spacetimes.

Notice that we have absorbed the deviation function $f$ into the Mino time, leaving the null geodesic equations unaffected. Similarly, the metric function $B$ can also be eliminated from the null geodesic equation through a change of the radial coordinate since it only appears in the expression for the radial coordinate velocity. Therefore, photon orbits in JP spacetimes are determined solely by the metric functions $N$ and $F$. As we will see below, neither $f$ nor $B$ appear in the expressions for any of the critical parameters. We will also show that a measurement of lensing parameters can be used to constrain these two metric functions. To constrain the remaining two functions, observations that do not rely on null geodesics may be required.

The full trajectory of arbitrary photons can now be written analytically in integral form, in terms of elliptic integrals, using the tangent given in Eq.~(\ref{eq:NG_Tangent}). 
More specifically, the total change in the coordinate time $\slashed{\Delta}t$, the azimuthal angle $\slashed{\Delta}\varphi$, the coordinate radius $\slashed{\Delta}r$, and the polar angle $\slashed{\Delta}\vartheta$ along an arbitrary null geodesic can all be obtained analytically. Since the latter pair of quantities are identically equal to the total change in the Mino time $\slashed{\Delta}\lambda_{\rm m}$, an arbitrary null geodesic connecting the source $x^{\mu}_{\rm s}$ to the observer $x^{\mu}_{\rm o}$ is completely characterized as \cite{Gralla:2019drh, Gralla:2019drh}
\begin{align} \label{eq:Elapsed_Quantities}
\begin{alignedat}{2}
\slashed{\Delta}t =&\ t_{\mathrm{o}} - t_{\mathrm{s}} 
&&=\ I_t + a^2 G_t\,, \\
\slashed{\Delta}\varphi =&\ \varphi_{\mathrm{o}} - \varphi_{\mathrm{s}} &&=\ I_\varphi + \xi G_\varphi\,,\\
\slashed{\Delta}\lambda_{\rm m} =&\ \lambda_{{\rm m; o}} - \lambda_{{\rm m; s}} 
&&=\ I_r = G_\vartheta\,,     
\end{alignedat}
\end{align}
where the subscripts $``\mathrm{o}"$ and $``\mathrm{s}"$ denote values of the various quantities at the observer and the source locations, respectively, with
\begin{align} 
I_t =&\ \fint_{r_{\mathrm{s}}}^{r_{\mathrm{o}}} \frac{\mathscr{T}_r(r)}{\pm_r\sqrt{\mathscr{R}(r)}}\mathrm{d}r\,;\,
G_t =\ \fint_{\vartheta_{\mathrm{s}}}^{\vartheta_{\mathrm{o}}} \frac{\cos^2{\vartheta}}{\pm_\vartheta\sqrt{\Theta(\vartheta)}}\mathrm{d}\vartheta\,,
\label{eq:Geodesic_Integrals1}
\\
I_r =&\ \fint_{r_{\mathrm{s}}}^{r_{\mathrm{o}}} \frac{\mathrm{d}r}{\pm_r\sqrt{\mathscr{R}(r)}}\,;\,\quad 
G_\vartheta =\ \fint_{\vartheta_{\mathrm{s}}}^{\vartheta_{\mathrm{o}}} \frac{\mathrm{d}\vartheta}{\pm_\vartheta\sqrt{\Theta(\vartheta)}}\,,
\label{eq:Geodesic_Integrals2}\\
I_\varphi =&\ \fint_{r_{\mathrm{s}}}^{r_{\mathrm{o}}} \frac{\Phi_r(r)}{\pm_r\sqrt{\mathscr{R}(r)}}\mathrm{d}r\,;\,
G_\varphi =\ \fint_{\vartheta_{\mathrm{s}}}^{\vartheta_{\mathrm{o}}} \frac{\csc^2{\vartheta}}{\pm_\vartheta\sqrt{\Theta(\vartheta)}}\mathrm{d}\vartheta\,.\label{eq:Geodesic_Integrals3}
\end{align}
The slash denotes that all of the above are path-dependent integrals either over the radial coordinate ($I_i$) or over the polar coordinate ($G_i$). 
As we will see below, all critical parameters are expressed in term of the elliptic functions from the $\vartheta$-integrals in Eqs.~(\ref{eq:Geodesic_Integrals1})-(\ref{eq:Geodesic_Integrals3}). 
Notably, the non-Kerr deviation functions appear exclusively in the radial integrals, while the angular integrals remain unaffected by these deviations, matching those of a Kerr BH. 

It is worthwhile to mention here that the change in the azimuthal angle $\slashed{\Delta}\varphi$ and the elapsed time $\slashed{\Delta}t$ are completely independent of the source and observer azimuthal coordinates $\varphi_{\rm s}$ and $\varphi_{\rm o}$ and also of their coordinate times $t_{\rm s}$ and $t_{\rm o}$. This invariance arises due to the axisymmetry and stationarity of the JP spacetime. For the Kerr limit of the JP metric, the integral geodesic equations in (\ref{eq:NG_Tangent}) match those in Eqs.~(7a)-(8f) of Ref.~\cite{Gralla:2019drh}.

\subsection{Null Geodesic Turning Points}\label{Sec-2b}
The path-dependent integrals above, Eqs.~(\ref{eq:Geodesic_Integrals1})-(\ref{eq:Geodesic_Integrals3}), that describe arbitrary photon trajectories can split into path-independent pieces \citep[see, e.g., Refs.][]{Shaikh+2019b, Gralla:2019drh, Gralla:2019drh, Kocherlakota+2024a}. The radial and angular potentials, $\mathscr{R}$ and $\Theta$ \eqref{eq:NG_Tangent}, which appear in these equations, remain positive everywhere except at the radial and polar ``turning points'' respectively, where the sign of the radial and polar coordinate velocities respectively reverse. Thus, over any segment of the orbit that is devoid of radial turning points, the integrands in $I_i$ do not change sign.

In particular, from the radial energy equation, $\dot{r}^2 = \mathscr{R}(r)$, it is clear that radial turning points occur at locations where $\mathscr{R} = 0$. Since we are interested in image formation on the screen of an asymptotic observer, we will only need to account for photon trajectories that contain at most one radial turning point outside the event horizon. 
Photons that are initially radially-ingoing ($-_r$) can only reach the observer if they meet a turning point whereas photons that are initially radially-outgoing ($+_r$) must not hit a turning point. The radial turning point is given by the largest real root of
\begin{equation} 
\label{eq:rTurning}
r - F a \xi - N\mathscr{I} = 0\,.
\end{equation}
Similarly, the polar turning points occur at $\vartheta = \vartheta_\pm$ where, with $u=\cos^2{\vartheta}$, the angular potential vanishes \cite{Gralla:2019drh, Gralla:2019drh}
\begin{equation} \label{eq:Polar_TPs}
(1-u)\Theta(u) = \eta - a^2u^2 + (a^2 - \eta - \xi^2)u = 0\,.
\end{equation}
The quadratic equation above has solutions
\begin{equation}\label{eq:u_sol}
u_\pm = \Delta_\vartheta \pm \sqrt{\Delta_\vartheta^2 + \frac{\eta}{a^2}}\,;\quad 
\Delta_\vartheta = \frac{1}{2}\left[1 - \frac{\eta+\xi^2}{a^2}\right].
\end{equation}
This equation also has a trivial solution $u=1$, which corresponds to the spherical coordinate singularity at $\vartheta=0,\pi$.
For positive $\eta$, it is clear that only $u_+ > 0$, which yields the polar turning points $\vartheta_\pm $,
\begin{equation}\label{eq:thetaTurning}
\vartheta_\pm = \arccos{\left(\mp\sqrt{u_+}\right)}\,.
\end{equation}
Solving Eqs.~(\ref{eq:rTurning}) and (\ref{eq:thetaTurning}) allows to classify photon trajectories in the phase space of $(\xi, \eta)$ based on the number and location of turning points \cite{Gralla:2019drh,Gralla:2019ceu}. A few noteworthy examples are as follows. Firstly, only those photons can reach an observer at $\vartheta_{\rm o}$ from a source at $\vartheta_{\rm s}$ for which $\vartheta_{\rm s}, \vartheta_{\rm o} \in [\vartheta_-, \vartheta_+]$, independent of the radial turning point. Furthermore, the polar turning points in Eq.~(\ref{eq:thetaTurning}) implies that no photon can reach the pole at $\vartheta=0$ or $\vartheta=\pi$ unless $\xi=0$. For $\xi\neq 0$, the polar motion is confined to oscillations between $\vartheta_{\pm}$ and $\vartheta_{\mp}$. Photons with $\eta > 0$,  referred to as following ``ordinary'' or ``oscillatory'' geodesics \cite{Gralla:2019drh, Gralla:2019drh}, oscillate between $\vartheta_-$ and $\vartheta_+$, above and below the equatorial plane, crossing the equatorial plane each time. The polar turning points are symmetric in each hemisphere: $0 \leq \vartheta_- \leq \frac{\pi}{2}$ and $\frac{\pi}{2} \leq \vartheta_+ \leq \pi$, with $\vartheta_+ = \pi - \vartheta_-$. 
This later feature might not be true for the spacetimes with broken $\textbf{Z}_2$ equatorial plane symmetry \cite{Chen:2020aix}. Photon with $\eta=0$ experience no polar turning points and follow circular orbits confined to the equatorial plane -- a type of ordinary geodesic.

In contrast, photons with $\eta < 0$, known as following ``vortical'' geodesics, have four polar turning points, two in each hemisphere. They exhibit two distinct motions, each oscillating between two turning points within a single hemisphere. Vortical geodesics do not intersect the equatorial plane and remain entirely within one hemisphere. 
For a source at the equatorial plane, photon trajectories connecting it to an observer are necessarily, therefore, always of the ordinary type. Notably, only ordinary geodesics contribute to photon ring formation, as vortical geodesics map to regions inside the critical curve on the image screen and do not affect the formation of the photon ring outside it \cite{Gralla:2019drh}.

Another important class of photon trajectories are those that stays on a constant $\vartheta$ surface, where both the polar angular velocity and acceleration vanish, i.e., $\dot{\vartheta}=\ddot{\vartheta}=0$. These planar geodesics with $(\xi, \eta)=(a\sin^2\vartheta, -a^2\cos^2\vartheta)$, known as principal null geodesics (PNG), are characterized by $\mathscr{I} = 0$ or $\eta < 0$ and trace the surface of a cone centered on the BH. These distinct types of photon trajectories will become relevant in the following subsections as we discuss the various limiting cases of the observer's inclination angle. 

\subsection{Spherical Null Geodesics}
\label{Sec-2c}

Due to strong gravitational lensing close to compact objects, photon orbits can become bound,%
\footnote{A distinct class of bound photon orbits, where photons oscillate between two distinct radii, may also exist. This occurs when a non-spinning BH has multiple, spatially-separated unstable and stable photon spheres in its exterior. It is possible also for a spinning BH spacetime to possess multiple photon shells. Here we are concerned with non-Kerr BHs (\ref{eq:JP_Metric}) that possess a single photon shell in its exterior containing bound, unstable SNGs.} %
which are of central interest here. Such a bound, spherical null geodesic (SNG) remains on a sphere of constant BL coordinate radius, $r_{\rm p}$, and, naturally, has vanishing radial coordinate velocity $\dot{r}$ and acceleration $\ddot{r}$, or, equivalently, $\mathscr{R} = 0$ and $\partial_r\mathscr{R} = 0$. These are the fixed points or critical points of the radial null geodesic flow equations. These pair of equations uniquely determine the critical values of conserved quantities or the impact parameters for an SNG with radius $r_{\rm p}$ as being,
\begin{align} \label{eq:SNG_Impact_Parameters}
\xi_{\mathrm{p}} = \left.\frac{1}{a}\frac{N - r\partial_r N}{N\partial_rF - F \partial_r N}\right|_{r=r_{\rm p}};\ \ 
\mathscr{I}_{\mathrm{p}} = -\left.\frac{F - r\partial_r F}{N\partial_rF- F \partial_r N}\right|_{r=r_{\rm p}}\,.
\end{align}
Henceforth, any quantity with the subscript ``${\rm p}$" is evaluated at an SNG of radius $r_{\rm p}$ with the impact parameters taking the corresponding critical values \eqref{eq:SNG_Impact_Parameters}. As alluded to above, only two deviation functions, $N$ and $F$, determine the critical values of impact parameters. These impact parameters describe both the prograde (co-rotating) and retrograde (counter-rotating) SNGs, distinguished by the sign of $\xi_{\rm p}$.

The region in the bulk of space permitting such SNGs, called the photon shell, comprises a set of spheres each of which permits such bound orbits that neither fall into BH nor escape to the infinity. The photon shell is located by finding the region in which the polar coordinate velocities $\dot{\vartheta}$ for SNGs are real-valued, i.e., with 
\begin{align} \label{eq:SNG_Theta}
\Theta_{\mathrm{p}}(\vartheta) := \mathscr{I}^2_{\mathrm{p}} - \left[\xi_{\mathrm{p}}\csc{\vartheta} - a\sin{\vartheta}\right]^2\,,
\end{align}
the photon shell lies between $r_{\mathrm{p}}^- \leq r_{\rm p} \leq r_{\mathrm{p}}^+$ such that $\Theta_{\mathrm{p}}(\pi/2)=\eta_{\mathrm{p}}=0$. The equatorial circular photon orbit of radius $r=r_{\mathrm{p}}^+$ is a retrograde orbit ($\xi_{\mathrm{p}} < 0$) whereas the one of radius $r=r_{\mathrm{p}}^-$ is a prograde orbit with $\xi_{\mathrm{p}} > a$. These two radii, $r=r_{\mathrm{p}}^{\pm}$, can be obtained as the largest roots respectively of the two equations
\begin{equation} \label{eq:NG_r_Eq}
\mp a(F - r\partial_r F) + (N - r\partial_r N) - a^2(N\partial_rF - F \partial_r N) = 0\,.
\end{equation}
For a general spinning BH, the photon shell is thickest at the equatorial plane, bounded by $r_{\rm p}^{\pm}$, and becomes vanishingly thin at the poles, with a radius of $r_{\rm p}^0$ with $\xi(r_{\rm p}^0)=0$. 
In contrast, for a non-spinning BH, Eq.~(\ref{eq:NG_r_Eq}) simplifies to $N - r \partial_r N = 0$, resulting in a degenerate photon sphere \cite{Salehi+2023}. Clearly, photons with $\eta_{\rm p}<0$ have $\Theta(\pi/2)<0$, thus they cannot move along SNG. Therefore, bound SNG in the photon shell are necessarily ``ordinary or oscillatory" orbits and not ``vortical" orbits.

\subsection{Image Plane Coordinates and the Shadow}\label{Sec-2d}
The Cartesian ``Bardeen" coordinates, $(\alpha, \beta)$, at which a photon appears on the screen of an observer are determined by its $4-$momentum, $k_\mu$, as well as the inclination or polar coordinate of the observer, $\vartheta=\mathscr{i}$, as \cite{Bardeen1973}
\begin{align} 
\begin{alignedat}{2}
\alpha =
    &\left .-r\frac{k_{(\varphi)}}{k_{(r)}}\right\rvert_{(r_{\rm o}, \mathscr{i})} 
        &&=\ -\xi\csc{\mathscr{i}}\,, \\
\beta =
    &\left .r \frac{k_{(\vartheta)}}{k_{(r)}}\right\rvert_{(r_{\rm o}, \mathscr{i})}  
        &&=\ \pm_\vartheta\sqrt{\Theta(\mathscr{i})}\,.
\end{alignedat}
\label{eq:Screen_Coords}
\end{align}
In the above, $r_{\mathrm{o}}$ represents the radial position of the observer, and in the writing the latter pair of equalities, we have assumed that the observer is far from the BH, $r_{\mathrm{o}} \gg M$. Furthermore, the bracketed subscripts ($k_{(a)}$) are used to denote that these are the components of the photon $4-$momentum projected onto an orthonormal tetrad adapted to the (static) observer, $\{e_{(a)}^\mu\}$, i.e., $k_{(a)} = k_\mu e_{(a)}^\mu$. Finally, the sign $\pm_\vartheta$ corresponds to the sign of the photon polar angular velocity, $k^\vartheta$, at the location of the observer.

We will work primarily with the image plane polar coordinates, ($\rho, \psi$), defined as
\begin{align}\label{eq:rho and psi}
\rho =&\ \sqrt{\alpha^2 + \beta^2} = \sqrt{\mathscr{I}^2 + 2a\xi - a^2\sin^2{\mathscr{i}}}\,,\\
\psi =&\ \arctan{\left(\frac{\beta}{\alpha}\right)}\,. \nonumber
\end{align}
SNG with conserved quantities $(\xi_{\mathrm{p}}, \mathscr{I}_{\mathrm{p}})$ can be used to describe a closed curve on the image plane, $\rho_{\mathrm{p}}(\psi_{\mathrm{p}})$. 
This is called the shadow boundary curve or the critical curve, which is symmetric about the $\alpha-$axis, and is independent of the metric functions $f$ and $B$. The critical curve is the gravitationally-lensed projection of the photon shell on the image plane. As the screen's polar angle $\psi$ changes along the critical curve, one moves across different photon spheres within the photon shell; one can invert Eq.~(\ref{eq:rho and psi}) to get $r_{\rm p}(\psi)$.

The condition $\Theta(\mathscr{i}) \geq 0$ determines the portion of the photon shell that contributes to the critical curve, for any given inclination angle. A polar observer, $\mathscr{i}=\{0,\pi\}$, accesses only a single photon sphere $r=r_{\rm p}^0$, and the shadow boundary curve is a perfect circle with radius $\rho_{\mathrm{p}}(\psi_{\mathrm{p}}) = \mathscr{I}_{\mathrm{p}}^0$. Only the equatorial observer sees the entire photon shell, with $\psi=0$ corresponding to the equatorial retrograde orbit and $\psi=\pi$ corresponding to the prograde one. For further discussion on null geodesics and various spherical photon orbits in Kerr spacetime, we direct the reader to see Refs. \cite{Teo2003, Gralla:2019xty, Gralla:2019drh}. 

\subsection{The Photon Ring}
\label{Sec-2e}
Photons on the SNGs remain trapped in bound orbits around the BH, and do not actually reach the critical curve on the image plane. Instead, what is observable as a proxy of critical curve are photon subrings, formed by ``nearly bound" photons that closely trace the photon shell.

Due to strong gravitational lensing, infinitely many light rays connect a source and an observer, located at arbitrary fixed spatial positions outside a BH. These are photons that execute a different number of orbits in the near-field region around the BH. Since photons that execute a larger number of orbits spend a longer amount of time close to the BH before reaching the observer, we can order these different photon orbits based on their respective total elapsed Mino times $\slashed{\Delta}\lambda_{\mathrm{m}} = G_{\vartheta}$ \eqref{eq:Elapsed_Quantities} between emission and observation, as described in Refs. \cite{Zhou:2024dbc} (see also Refs. \cite{Kapec+2020, Gralla:2019drh} for alternative ordering schemes). 

In terms of the Mino time elapsed over a half-polar orbit, i.e., the Mino time taken for a photon to travel from one polar turning point to the other,
\begin{equation} \label{eq:MinoTime}
\hat{G}_{\vartheta} := \int_{\vartheta_-}^{\vartheta_+}\frac{1}{\sqrt{\Theta(\theta)}}\mathrm{d}\vartheta=\frac{2}{a\sqrt{-u_-}}K\left(\frac{u_+}{u_-}\right)\,,
\end{equation}
where $K(k)$ is the complete elliptic integral of the first kind, 
\begin{equation}
K(k) = \int_0^{\pi/2}\frac{\mathrm{d}\vartheta}{\sqrt{1-k\sin^2{\vartheta}}}\,,
\end{equation}
we can define the half winding index of an arbitrary photon orbit, i.e., the (fractional) number of half-polar orbits executed close to the photon shell, as
\begin{equation}
w_{1/2}=\frac{G_{\vartheta}}{\hat{G}_{\vartheta}}\,.
\end{equation}
The order of the photon orbit $n$ is then defined simply as \cite{Zhou:2024dbc}
\begin{equation}
n = \left \lfloor{w_{1/2}}\right \rfloor.
\end{equation}
Photons with larger path lengths have larger Mino times, and, thus, larger $n$. Ref. \cite{Zhou:2024dbc} also discusses the important possibility of specific spatial configurations of source and emitter causing multiple photon orbits connecting them to have the same image order or ``level,'' $n$. In such cases, the orbits are further sub-classified by their half winding index, $w_{1/2}$. 

Please note that the labels for the half winding index and the image order have varied across recent papers. We adopt the convention above simply to ensure that $n$ corresponds to an \textit{integer-valued} quantity that solely determines the order of the image. This convention enables us to write equations such as Eqs.~(\ref{eq:Eq_Pole_Scaling_Relations}) and (\ref{eq:Radial_Dev_Sol_Polar_Angle1}) in their now familiar form \cite{Johnson:2019ljv}. This convention also matches that of Ref. \cite{Kocherlakota+2024a}, where the simpler problem of ordering images in arbitrary spherically-symmetric spacetimes was revisited.

Image formation in the photon ring is governed by nearly-bound photon orbits, which have essentially the same impact parameters as the bound SNGs. Thus, at a particular polar angle $\psi$ on the image plane, the impact parameters of the nearly-bound SNGs are given by those of the SNG at a particular radius $r_{\mathrm{p}}(\psi)$, described in the previous section. Therefore, the Mino half-period, $\hat{G}_\vartheta$, for a nearly-bound photon orbit varies over the screen angle $\psi$ -- monotonically increasing with $\psi$ from $\psi=0$ to $\psi=\pi$. As we radially move across the SNGs within the photon shell from $r_{\rm p}^+$ to  $r_{\rm p}^-$ the $\hat{G}_{\vartheta}$ monotonically increases. In particular, for retrograde nearly-bound photon orbits with turning points $\vartheta_-$ and $\vartheta_+$ located closer to each other have shorter Mino half-periods. Intriguingly, we note that photons with the same impact parameters ($\xi, \mathscr{I}$) in different JP spacetimes (including Kerr) with identical spins have identical Mino half-periods.

As a result of the above, a point source near the BH produces infinitely many images, indexed by the number $n$. The zeroth-order (or direct) image, $n=0$, is formed by the photon that has the shortest path-length, the first-order (or first-indirect) image, $n=1$, has the second shortest path-length, etc. For an extended source of emission, these higher-order images manifest as a discrete sequence of bright, narrow, nearly circular photon subrings indexed by $n$, collectively referred to as ``the photon ring" \cite{Gralla:2019xty}. These subrings are stacked on top of the direct image, creating a ``wedding cake" structure \cite{Gralla:2019xty,Vincent:2022fwj}. Each $n^{\mathrm{th}}$-order photon subring represents the collection of all order-$n$ images from every point source in the emission region. These rings converge exponentially, with $n$, toward the BH's critical curve.

Photon rings are routinely observed in general relativistic magnetohydrodynamic simulations of accretion flows around Kerr BHs, manifesting as sharp, periodic features in both images and the visibility domain. In the visibility domain, photon rings appear as a ``ringing" pattern \cite{Johnson:2019ljv,Gralla:2020yvo,Paugnat+2022,Cardenas-Avendano:2023dzo,Palumbo:2023auc}. As the photon subring order $n$ increases, the influence of the accretion physics diminishes, with the $n = 2$ subring already being largely independent of astrophysical models and closely aligned with the critical curve \cite{Kocherlakota+2024a,Gralla:2020yvo}. The higher-order subrings ($n \gtrsim 2$) exhibit a remarkable universality in shape \cite{Gralla:2020yvo,Gralla:2020nwp} and follow simple scaling relations governed by three astrophysics-independent critical parameters: $\gamma_{\rm p}$, $\tau_{\rm p}$, and $\delta_{\rm p}$. These critical parameters -- the Lyapunov exponent $\gamma_{\rm p}$, the delay time $\tau_{\mathrm{p}}$, and the rotation parameter $\delta_{\rm p}$ -- control the radial demagnification, coordinate time delay, and rotation of successive photon subrings on the image plane, respectively \cite{Gralla:2019drh}. For an equatorial disk of emission viewed by a polar observer, if we denote the radius of the outer edge of its $n^{\mathrm{th}}-$order image by $\rho_n$, the time at which it appears on the screen by $t_n$, and the polar angle at which it appears by $\psi_n$, then these scaling relations are given as \cite{Gralla:2019drh}:
\begin{equation} \label{eq:Eq_Pole_Scaling_Relations}
\frac{\mathscr{\rho}_{n+1}-\mathscr{\rho}_{\rm p}}{\mathscr{\rho}_{n}-\mathscr{\rho}_{\rm p}}\approx e^{-\gamma_{\rm p}},\;\; 
t_{n+1}-t_{n}\approx\tau_{\rm p},\;\;
\psi_{n+1}-\psi_{n}\approx \delta_{\rm p}\,,
\end{equation}
whereas, for general source and observer orientations, the relations above receive corrections [see Ref.~\cite{Gralla:2019drh} for further details.] 

Therefore, the detection of the $n = 2$ photon subring is particularly valuable for disentangling gravitational and astrophysical effects and for studying gravity in the strong-field regime. The prospects of observing up to the $n = 2$ photon subring of the M87$^*$ BH using ground-space VLBI in the near future have been put forward \cite{Gralla:2020srx,Lupsasca:2024xhq}.

\subsection{Photon Ring Critical Parameters}\label{Sec-2f}
Following Ref.~\cite{Gralla:2019drh}, we now determine the three critical photon ring parameters ${\gamma, \tau, \delta}$ for this broad class of stationary and axisymmetric, integrable spacetimes \eqref{eq:JP_Metric}. The critical parameters, which govern the universal features of the photon subring as described above, are defined over a polar half-orbit (from $\vartheta_{\pm}$ to $\vartheta_{\mp}$) of a particular bound SNG at $r = r_{\mathrm{p}}$.  
The critical parameters depend solely on the spacetime geometry and the observer's inclination, and are independent, in particular, of the accretion physics around astrophysical BHs. Therefore, a measurement of the critical parameters offers a powerful tool for experimental tests of gravity.

These critical parameters additionally depend on the SNG radius $r_{\rm p}$, and therefore, for a spinning BH observed at inclinations $\mathscr{i}\neq 0, \pi$, they vary with the screen angle $\psi_{\rm p}$ along the critical curve. However, for a polar observer, where the critical curve is circular, these parameters remain constant along the curve. 

We consider a light ray with critical values for the impact parameters ($\xi = \xi_{\mathrm{p}}, \mathscr{I} = \mathscr{I}_{\mathrm{p}}$) and tracing an orbit close to that of the SNG with the same impact parameters. If the radius of the latter is denoted by $r(\lambda_{\rm m})=r_{\mathrm{p}}$, then we can denote that of the former as $r(\lambda_{\rm m}) = r_{\mathrm{p}} + \delta r(\lambda_{\rm m})$, with $|\delta r(\lambda_{\rm m})| \ll r_{\rm p}$. For such a photon, as long as $|\delta r(\lambda_{\rm m})| \ll r_{\rm p}$, we can write, from the radial energy equation \cite{Johnson:2019ljv},
\begin{align} \label{eq:Radial_Dev_Eq_Mino}
\left(\dot{\delta r}\right)^2 \approx&\ \mathscr{R}_{\mathrm{p}} + \left[\partial_r\mathscr{R}_{\mathrm{p}}\right] (\delta r) + \left[\frac{\partial_r^2\mathscr{R}_{\mathrm{p}}}{2}\right] (\delta r)^2\,, \nonumber \\
\approx&\ \kappa_{\mathrm{p}}^2(\delta r)^2\,;\quad 
\kappa_\mathrm{p} = \sqrt{\frac{\partial_r^2\mathscr{R}_{\mathrm{p}}}{2}}\,,
\end{align}
where, in writing the latter equation, we have used the fact that at the bound photon orbit, $\mathscr{R}_{\mathrm{p}}=\partial_r \mathscr{R}_{\mathrm{p}}=0$. 
Equation (\ref{eq:Radial_Dev_Eq_Mino}) is the linearized radial geodesic equation, and can be solved analytically to obtain 
\begin{align} \label{eq:Radial_Dev_Sol_Mino}
\log{\left[\frac{\delta r(\lambda_{\mathrm{m}})}{\delta r_{\mathrm{i}}}\right]} \approx \pm\kappa_{\mathrm{p}}\lambda_{\rm m}\quad 
\mathrm{or}\quad
\delta r(\lambda_{\mathrm{m}})=\delta r_{\mathrm{i}}\,e^{\pm\kappa_{\mathrm{p}}\lambda_{\rm m}}.
\end{align}
In the above, $\delta r_{\mathrm{i}}$ corresponds to the initial radial deviations of the photon orbit from the nearest SNG. For photons on radially-perturbed orbits that are moving away from (towards) the SNG, one picks the positive (negative) sign in the equations above.

The radial deviation of a photon that is initially moving away from a nearby SNG grows exponentially, at a rate determined by the null geodesic phase space Lyapunov exponent $\kappa_{\mathrm{p}}$ \cite{Deich:2023oox}. This exponent is not a reparametrization-invariant quantity since it is defined relative to an affine parameter ($\lambda_{\rm m}$). Nevertheless, it determines the radial stability of the SNG on the photon sphere at $r=r_{\mathrm{p}}$ as being unstable (stable) if $\kappa_{\mathrm{p}}$ is real (imaginary). Larger Lyapunov exponents imply that null geodesics close-by (in phase space) to an SNG diverge more quickly in affine time from it. For the Schwarzschild metric in particular, $\kappa_{\mathrm{p}} = \sqrt{27}M$.  The difference from Eq.~(2.7) of Ref.~\cite{Kocherlakota+2024b} arises due to the reparametrization in Eq.~(\ref{eq:Mino_Time}).

We can rewrite the radial deviation equation \eqref{eq:Radial_Dev_Eq_Mino} in terms of the polar angle along the orbit as
\begin{equation} \label{eq:Radial_Dev_Eq_Polar_Angle}
\frac{\dot{\delta r}}{\dot{\vartheta}} \approx \pm \frac{\kappa_{\mathrm{p}}\cdot\delta r}{\sqrt{\Theta_{\mathrm{p}}}}\,.
\end{equation}
If after $n$ half-polar orbits (each either from $\vartheta_- \rightarrow \vartheta_+$ or from $\vartheta_+ \rightarrow \vartheta_-$), the radial position of the photon is $\delta r(\lambda_{\rm m} = \lambda_{\rm m;f}) = \delta r_{\rm f}$, then from the previous equation we can find
\begin{equation}
 \begin{aligned} \label{eq:Radial_Dev_Sol_Polar_Angle1}
\log{\left[\frac{\delta r_{\mathrm{f}}}{\delta r_{\mathrm{i}}}\right]} 
\approx & \pm n\,\kappa_{\mathrm{p}}\int_{\vartheta_-}^{\vartheta_+} \frac{\mathrm{d}\vartheta}{\sqrt{\Theta_{\mathrm{p}}}} 
= \pm n \kappa_{\mathrm{p}} \hat{G}_\vartheta
=: \pm n \gamma_{\rm p}\,\\
\delta r_{\mathrm{f}} = & \; \delta r_{\mathrm{i}}\, e^{\pm n \gamma_{\rm p}}.   
\end{aligned}
\end{equation}

In the above, we have introduced the ``scaling'' or ``de-magnification'' Lyapunov exponent $\gamma_{\mathrm{p}}$ as,
\begin{equation} \label{eq:gamma_p_Def}
\gamma_{\mathrm{p}} := \kappa_{\mathrm{p}}\hat{G}_\vartheta\ 
= \sqrt{\frac{\partial_r^2\mathscr{R}_{\mathrm{p}}}{2}} \frac{2}{a\sqrt{-u_-}}K\left(\frac{u_+}{u_-}\right).
\end{equation}
While the phase space Lyapunov exponent $\kappa_{\mathrm{p}}$ measures the variation in the radial deviation $\delta r$ with Mino time $\lambda_{\rm m}$, the demagnification Lyapunov exponent $\gamma_{\mathrm{p}}$ measures the same relative to changes in the polar angle $\vartheta$ over a polar half-orbit. Since this means that $\gamma_{\mathrm{p}}$ can be defined unambiguously, independent of the parametrization of the orbit or the coordinate system used, it is therefore a potentially measurable quantity \cite{Johnson:2019ljv}.

The Lyapunov exponent, $\gamma_{\mathrm{p}}$, not only governs the radial trajectory of nearly-bound photons but also controls the width and flux of the resulting photon subrings on the image screen. While a light ray completing $n$ half-orbits in the optically thin emission region around a BH carries $n$ times more photons than one crossing the region only once, the transfer function becomes exponentially steeper with increasing $n$ \cite{Gralla:2019xty}. Consequently, although the local brightness of photon subrings increases linearly with $n$, the width and flux of each successive subring diminishes exponentially (slope of the transfer function), controlled by the demagnification rate set by $\gamma_{\mathrm{p}}$. 
We direct the reader to see Ref.~\cite{Gralla:2019drh} (see Eq.~(74) and Appendix B there) and Ref.~\cite{Johnson:2019ljv} for further instructive insights on how the demagnification Lyapunov exponent relates successive photon subrings, showing them to exponentially decrease in width and also exponentially limiting to the critical curve.

The coordinate time elapsed over one half-polar orbit, $\tau_{\mathrm{p}} := \slashed{\Delta}t_{\mathrm{p}}$, can be obtained using Eq.~(\ref{eq:Elapsed_Quantities}), as being 
\begin{align} \label{eq:tp_Def}
\tau_{\mathrm{p}} := & \ I_t(r_{\mathrm{p}}+\delta r_{\mathrm{i}}, r_{\mathrm{p}} + \delta r_{\mathrm{f}}) + a^2 \hat{G}_t\,, \nonumber \\
\approx&\ \mathscr{T}_{r; \mathrm{p}}\hat{G}_\vartheta + a^2\hat{G}_t\,, \nonumber \\
\tau_{\mathrm{p}} := &\ \left[\frac{r_{\mathrm{p}}}{N_{\mathrm{p}}}\mathscr{I}_{\mathrm{p}} + a\xi_{\mathrm{p}} - a^2\right]\hat{G}_\vartheta + a^2\hat{G}_t\,.
\end{align}
In the above, $\hat{G}_t$ is given as \cite{Kapec+2020, Gralla:2019drh}
\begin{equation} \label{eq:hat_G_t}
\hat{G}_t := \int_{\vartheta_-}^{\vartheta_+}\frac{\cos^2\vartheta}{\sqrt{\Theta(\theta)}}\mathrm{d}\vartheta = -\frac{4u_+}{a\sqrt{-u_-}}E'\left(\frac{u_+}{u_-}\right)\,,
\end{equation}
where $E'(k)$ denotes the derivative of the complete elliptic integral of the second kind w.r.t. the elliptic modulus $k$, i.e., $E'(k) := [E(k) - K(k)]/(2k)$. We adopt the convention
\begin{equation}
E(k) = \int_0^{\pi/2}\sqrt{1-k\sin^2{\vartheta}}\mathrm{d}\vartheta\,.
\end{equation}

Light rays forming successive photon subrings on the image screen differ in path length by an additional turning point or half-polar orbit around the BH.
Therefore, for sufficiently large ring order $n\geq 2$,
\footnote{Higher-order rings refer to photon subrings that are largely independent of the astrophysical accretion and emission model around the BH.} 
the time delay $\tau_{\rm p}$ represents the elapsed coordinate time over each half-polar orbit and serves as a measure of the delay in the appearance of successive subrings on the image screen. Note that light rays with fixed $(\xi, \mathscr{I})$ moving in Kerr and JP spacetimes, have identical elapsed Mino times $\hat{G}_{\vartheta}$ but different coordinate times $\tau_{\mathrm{p}}$ over one half-polar orbit.

Since the coordinate-time period of a half-polar orbit, $\tau_{\rm p}$, depends on the SNG radius $r_{\rm p}$, and for an extended emission source, a $n^{\rm th}$-order photon subring consists of all $n^{\rm th}$-order images from SNGs within the photon shell, the time delay consequently varies along the critical curve. This implies that, for inclinations $\mathscr{i} \neq 0,\pi$, different sections of the photon subring take different amounts of time to become visible, leading to a temporal variation across the subring structure. This time delay could be observed for time-varying emission sources around a BH, such as flares. 

In a spinning spacetime, an arbitrary null geodesic does not return to its initial azimuthal coordinate after completing one full polar orbit ($\vartheta_\mp \rightarrow \vartheta_\pm \rightarrow \vartheta_\mp$) in general. 
Resonant orbits, which return to the same azimuthal coordinate, after an integral number of complete polar-orbits have been discussed in Refs. \cite{Teo2003, Wong2021}. 
We can characterize the effect of frame-dragging or gravitomagnetism on photon orbits via the total change in the azimuthal angle $\varphi$ over a complete polar orbit for a nearly bound null geodesic, $2\slashed{\Delta}\varphi_{\mathrm{p}}$, which is given as 
\begin{align} \label{eq:varphi_p_Def}
\slashed{\Delta}\varphi_{\mathrm{p}} :=&\ I_\varphi(r_{\mathrm{p}}+\delta r_{\mathrm{i}}, r_{\mathrm{p}}+\delta r_{\mathrm{f}}) + 2\xi_{\mathrm{p}}\hat{G}_\varphi\,, \nonumber \\
\approx&\ 2\Phi_{r; \mathrm{p}}\hat{G}_\vartheta + 2\xi_{\mathrm{p}}\hat{G}_\varphi\,, \nonumber \\
=&\ 2a\ \left[\frac{F_{\mathrm{p}}}{N_{\mathrm{p}}}\mathscr{I}_{\mathrm{p}} - 1\right]\hat{G}_\vartheta + 2\xi_{\mathrm{p}}\hat{G}_\varphi\,.
\end{align} 
In the above, $\hat{G}_\varphi$ is given as \cite{Kapec+2020, Gralla:2019drh}
\begin{equation}
\hat{G}_\varphi := \int_{\vartheta_-}^{\vartheta_+}\frac{\csc^2\vartheta}{\sqrt{\Theta(\theta)}}\mathrm{d}\vartheta = \frac{2}{a\sqrt{-u_-}}\Pi\left(u_+ \middle| \frac{u_+}{u_-}\right)\,,
\end{equation}
where $\Pi(n|k)$ is the complete elliptic integral of the third kind, with $n$ denoting the elliptic characteristic, and is given as
\begin{equation}
\Pi(n|k) = \int_0^{\pi/2}\frac{1}{(1-n\sin^2{\vartheta})\sqrt{1-k\sin^2{\vartheta}}}\mathrm{d}\vartheta\,.
\end{equation}

A SNG that reaches the north ($\vartheta = 0$) and south ($\vartheta = \pi$) poles is located at a radius $r=r_{\mathrm{p}}^0$. As discussed above, a photon on such an orbit has zero angular momentum, $\xi_{\mathrm{p}} = 0$. This ``polar'' SNG divides the photon shell into two pieces, (a.) spheres of radii $r > r_{\mathrm{p}}^0$ that are filled with counter-rotating photons possessing negative angular momenta ($\xi_{\mathrm{p}} < 0$) and (b.) spheres of radii $r < r_{\mathrm{p}}^0$ which contain co-rotating photons with positive angular momenta ($\xi_{\mathrm{p}} > 0$). On the image plane, the impact parameters of the polar SNGs correspond to the antipodal points on the $\beta-$axis where the critical curve intersects it. Counter-rotating photons appear on the positive$-\alpha$ segment of the critical curve whereas the co-rotating ones appear on the negative$-\alpha$ segment.

Due to the change in the sense of rotation of a counter-rotating bound photon relative to a co-rotating one, both on spheres of radii approximately $r\approx r_{\mathrm{p}}^0$, the elapsed azimuthal angle, $\slashed{\Delta}\varphi_{\rm p}$, jumps discontinuously from one orbit to another. Nevertheless, the difference between the two is precisely $4\pi$ \cite{Teo2003}. Each prograde (retrograde) orbit takes more (less) than one revolution in the azimuthal angle to complete one full polar-orbit and vice versa.

For the zero angular momentum bound orbit, we can find (cf. Eq.~(69) of Ref.~\cite{Gralla:2019drh})
 \begin{equation}
 \slashed{\Delta}\varphi_{\mathrm{p}} = a\at{\left[\frac{F_{\mathrm{p}}}{N_{\mathrm{p}}}\mathscr{I}_{\mathrm{p}} - 1\right]\frac{4}{\mathscr{I}_{\mathrm{p}}}K\left(\frac{a^2}{\mathscr{I}_{\mathrm{p}}^2}\right)}{r_{\mathrm{p}}=r_{\mathrm{p}}^0} + 2\pi
\end{equation}

For the aforementioned reason, following Ref.~\cite{Gralla:2019drh} (see Eq.~(48) there), we introduce a rotation parameter $\delta_{\mathrm{p}}$, which describes the change in the azimuthal angle over a half-polar orbit of a nearly-bound null geodesic, and which varies smoothly across the polar SNG radius, as
\begin{equation}
\label{eq:varphi_p_Def1}
 \delta_{\mathrm{p}} := a  \left[\frac{F_{\mathrm{p}}}{N_{\mathrm{p}}}\mathscr{I}_{\mathrm{p}} - 1\right]\hat{G}_\vartheta + \xi_{\mathrm{p}}\hat{G}_\varphi + 2 \pi H(r-r_{\rm p}^0)\,,
\end{equation} 
where $H(x)$ is the Heaviside (step) function. Interestingly, unlike $\gamma_{\rm p}$ and $\tau_{\rm p}$, a $\delta_p \neq \pi$ necessarily implies that the BH is spinning. By definition, $\delta_{\rm p}$ not only encodes the change in $\varphi$ accrued by a nearly-bound photon over each polar half-orbit but also determines the angular shift of successive photon subrings on the image screen, an effect which would be visible for non-axisymmetric emission sources. As discussed earlier, $\delta_{\rm p}$ varies with the screen polar angle $\psi$, indicating that at high inclinations, different sections of the photon subrings experience varying degrees of rotation.

In non-spinning spacetimes ($a=0$), we can write
\begin{align} \label{eq:Hatted_Integrals_a0}
\hat{G}_\vartheta = \int_{0}^{\pi} \frac{\mathrm{d}\vartheta}{\mathscr{I}_{\mathrm{p}}} = \frac{\pi}{\mathscr{I}_{\mathrm{p}}}\,; \
\hat{G}_t = \int_{0}^{\pi} \frac{\cos^2{\vartheta}}{\mathscr{I}_{\mathrm{p}}}\mathrm{d}\vartheta = \frac{\pi}{2\mathscr{I}_{\mathrm{p}}}\,, 
\end{align}
and the three critical parameters are then given as (see, e.g., Ref. \cite{Kocherlakota+2024b}),
\begin{align}
\gamma_{\mathrm{p}} = \frac{\pi\kappa_{\mathrm{p}}}{\mathscr{I}_{\mathrm{p}}}\,;\ 
\tau_{\mathrm{p}} = \pi \mathscr{I}_{\mathrm{p}}\,;\ 
\delta_{\mathrm{p}} = \pi\,. 
\end{align}
Thus, for non-spinning spacetimes, the time delay $\tau_{\rm p}$ in formation of successive photon subrings on the image plane provides an indirect measure of the BH shadow size $\mathscr{I}_{\rm p}$, and consecutive images of a single source appear on antipodal points on the image plane.

For a non-Kerr JP BH, the analytical expressions for the three critical parameters, controlling the demagnification, coordinate time delay, and the rotation of successive photon subrings, ($\gamma_{\rm p}, \tau_{\rm p}, \delta_{\rm p}$) are presented in Eqs.~(\ref{eq:gamma_p_Def}), (\ref{eq:tp_Def}), and (\ref{eq:varphi_p_Def1}), respectively. These depend on the BH mass, spin, deviation functions $F$ and $N$, and can be viewed as functions of the image plane polar angle $\psi$ of the critical curve.
\footnote{We note that while the metric function $B$ appears in the expression for $\gamma_{\mathrm{p}}$, it can be eliminated through a change of the radial variable, as discussed above. Essentially, such a change will simply remove $B$ from the radial potential in Eq.~(\ref{eq:NG_Tangent}), leaving $\kappa_{\mathrm{p}}$ and $\gamma_{\mathrm{p}}$ to be determined by $N$ and $F$.} 
Therefore, these are determined purely by the spacetime geometry and the observer's inclination angle, and are entirely independent of the details of the emission source. Furthermore, these expressions correctly reduce to those for the Kerr BH, as given in Ref.~\cite{Gralla:2019drh}.

Moreover, the presented framework applies broadly to any non-rotating or rotating BH metrics in general relativity or modified gravity theories that map to the JP metric (\ref{eq:JP_Metric}). The key criteria here are the photons following the null geodesics of metric (\ref{eq:JP_Metric}) and the separability of the null geodesic equation or the existence of a Killing tensor. 
An approach to calculate similar critical parameters in a non-separable BH spacetime is not yet clear.

\section{Observational Signatures for Polar Observer}
\label{Sec-3}

In this section, we examine the perspective of an observer positioned along the rotational axis of a central compact object. Previous studies have extensively explored the shadow size and Lyapunov exponent observed from a polar viewpoint in a general axisymmetric spacetime (see \cite{Salehi+2023}). Here, we extend this analysis to additional observables, including the time delay and azimuthal shift.

The shadow size $\rho_{\mathrm{p}}$, the demagnification parameter $\gamma_{\mathrm{p}}$, the delay time $\tau_{\mathrm{p}}$, and the rotation parameter $\delta_{\mathrm{p}}$ are all constants for the polar observer in a fixed spacetime. This is because only zero angular momentum geodesics, with turning points at $\vartheta_- = 0$ and $\vartheta_+ =\pi$, can reach the polar observer. Since the zero angular momentum SNG, in particular, orbits on a complete sphere of radius $r=r_{\mathrm{p}}^0$, we can obtain the critical quantities for a polar observer by fixing $r$ to this value in the general expressions obtained in the previous section. For simplicity, we will drop the subscript ``$\mathrm{p}$'' but keep ``$0$'' in the discussion below. The radius of the polar orbit $r_0$ and the polar shadow radius $\rho_0$ are given respectively as \cite{Salehi+2023}
\begin{equation}
\label{eq:r polar}
r_0 = \frac{N_0}{\partial_r N_0},\quad \rho_0 = \mathscr{I}_0 = \frac{1}{\partial_r N_0}.
\end{equation}
With the solutions to the turning point equation \eqref{eq:Polar_TPs},
\begin{equation} \label{eq:Polar_Turning_Points_Polar_Obs}
u_{+;0} = 1\,;\ u_{-;0} = -\eta_0/a^2\,,
\end{equation}
we can obtain the polar values of all the crtical parameters from the general expressions reported in the previous section. 

In particular, for a zero angular momentum SNG, we only need $\hat{G}_{\vartheta; 0}$ and $\hat{G}_{t; 0}$ to obtain the photon ring critical parameters. These are given as,
\footnote{Note that we have used standard negative-argument identities for complete elliptic functions to eliminate $\eta_0$ in favor of $\mathscr{I}_0$.}
\begin{equation}
\hat{G}_{\vartheta; 0} = \frac{2}{\mathscr{I}_0}K\left(\frac{a^2}{\mathscr{I}_0^2}\right)\,;\ 
\hat{G}_{t; 0} = \hat{G}_{\vartheta; 0} + \frac{4}{\mathscr{I}_0}E'\left(\frac{a^2}{\mathscr{I}_0^2}\right)\,,
\label{eq:polar G_theta and G_t}
\end{equation}
with which
\begin{equation}
\begin{aligned} \label{eq:Polar_Crit_Params}
\gamma_0 &=\ \frac{\kappa_0}{\mathscr{I}_0}2K\left(\frac{a^2}{\mathscr{I}_0^2}\right) = \left[\sqrt{-\frac{N^3\partial_r^2 N}{(\partial_rN)^2B^2}}\right]_0 2K\left(\frac{a^2}{\mathscr{I}_0^2}\right)\,,
\\
\tau_0 &=\ 2\mathscr{I}_0E\left(\frac{a^2}{\mathscr{I}_0^2}\right)\,, \\
\delta_0 &=\ a \left[\frac{F}{N} \mathscr{I} - 1\right]_0\frac{2}{\mathscr{I}_0}K\left(\frac{a^2}{\mathscr{I}_0^2}\right) + \pi\,,\\
&=a \left[\frac{F}{r}\mathscr{I}_0^2 - 1\right]_0\frac{2}{\mathscr{I}}K\left(\frac{a^2}{\mathscr{I}_0^2}\right) + \pi\,.
\end{aligned}
\end{equation}

In the following subsections, we will explore the observational signatures of some of these potential observables.

The location of the event horizon ($g^{rr} = 0$) and the shape of the ergosurface ($g_{tt}=0$) are determined by the metric functions \( N \) and \( F \). In particular, in spherically-symmetric spacetimes, $F=0$ \cite{Salehi+2023}. Therefore, a constraint of a non-zero value for \( F \) would indicate the existence of an ergoregion. The latter is a crucial ingredient in Penrose processes  \cite{Penrose1969}, i.e., classical processes through which energy can be extracted from BHs \cite{Lasota:2013kia}. The most astrophysically-relevant Penrose process is the Blandford-Znajek process, which is thought to power jets, i.e., relativistic outflows of magnetically-dominated plasmas \cite{Blandford+1977, Tchekhovskoy+2011, Lasota:2013kia}.

The polar time delay $\tau_0$ is entirely characterized by the spin $a$ and the shadow size $\mathscr{I}_0$. Since the latter is given by the first derivative of the metric function \( N \) at the critical photon radius, $\partial_rN_0$, or, equivalently, by the ratio $(N/r)_0$, the delay time remains uninformative about \( F \). 

\begin{figure}
    \centering
    \includegraphics[width= 1.1\columnwidth]{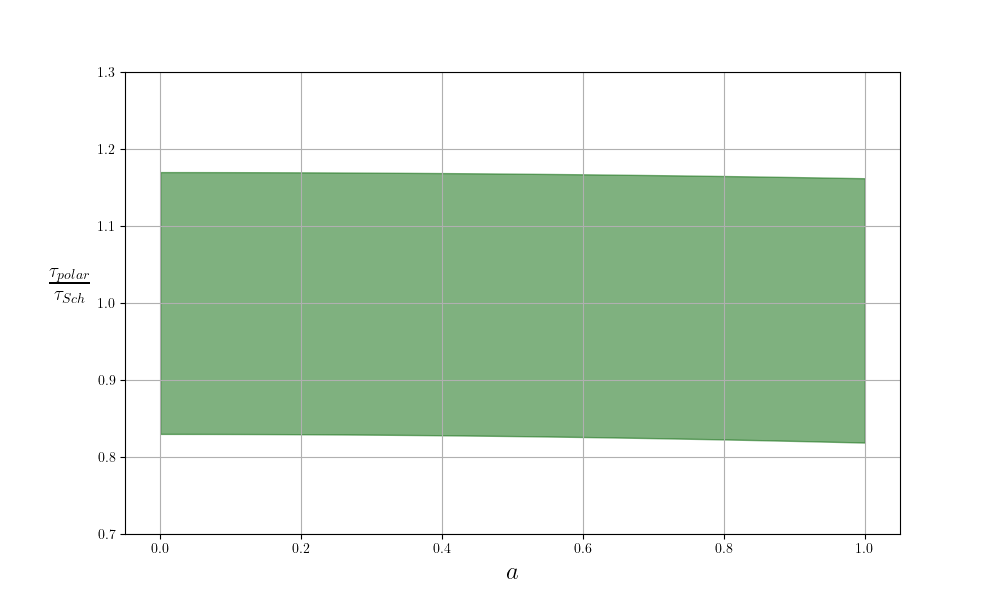}
    \caption{Shown in green is the region of combined values of time delay and BH spin that are consistent with the $1 \sigma$ 2017 EHT shadow size measurement of M87$^*$. We show the time delay normalized to its Schwarzschild value, $\sqrt{27}\pi M$, on the $y-$axis.}
    \label{fig:time_delay_polar}
\end{figure}

Currently, with the 2017 shadow size measurement of M87$^*$ reported by the EHT in hand, $\rho_0 \approx \sqrt{27}M(1+0.^{+0.17}_{-0.17})$ \cite{EHTC+2019f, Psaltis+2020, Kocherlakota+2021}, we can already predict time delay that we should expect to measure from future measurements, for all possible BH spins. This prediction is illustrated in Fig. \ref{fig:time_delay_polar}, where the y-axis represents the polar value of the time delay, normalized by the time delay in a Schwarzschild spacetime, \( \sqrt{27}\pi M \) \cite{Gralla:2020nwp}. The green bands in this plot represents the regions that are consistent with the aforementioned EHT shadow size measurement.

Finally, with improved shadow size measurements and potential future time delay measurements, the valid region in Fig. \ref{fig:time_delay_polar} would become more constrained, allowing for a possible measurement of the BH spin.

\begin{figure*}
    \centering
    \includegraphics[width= 2\columnwidth]{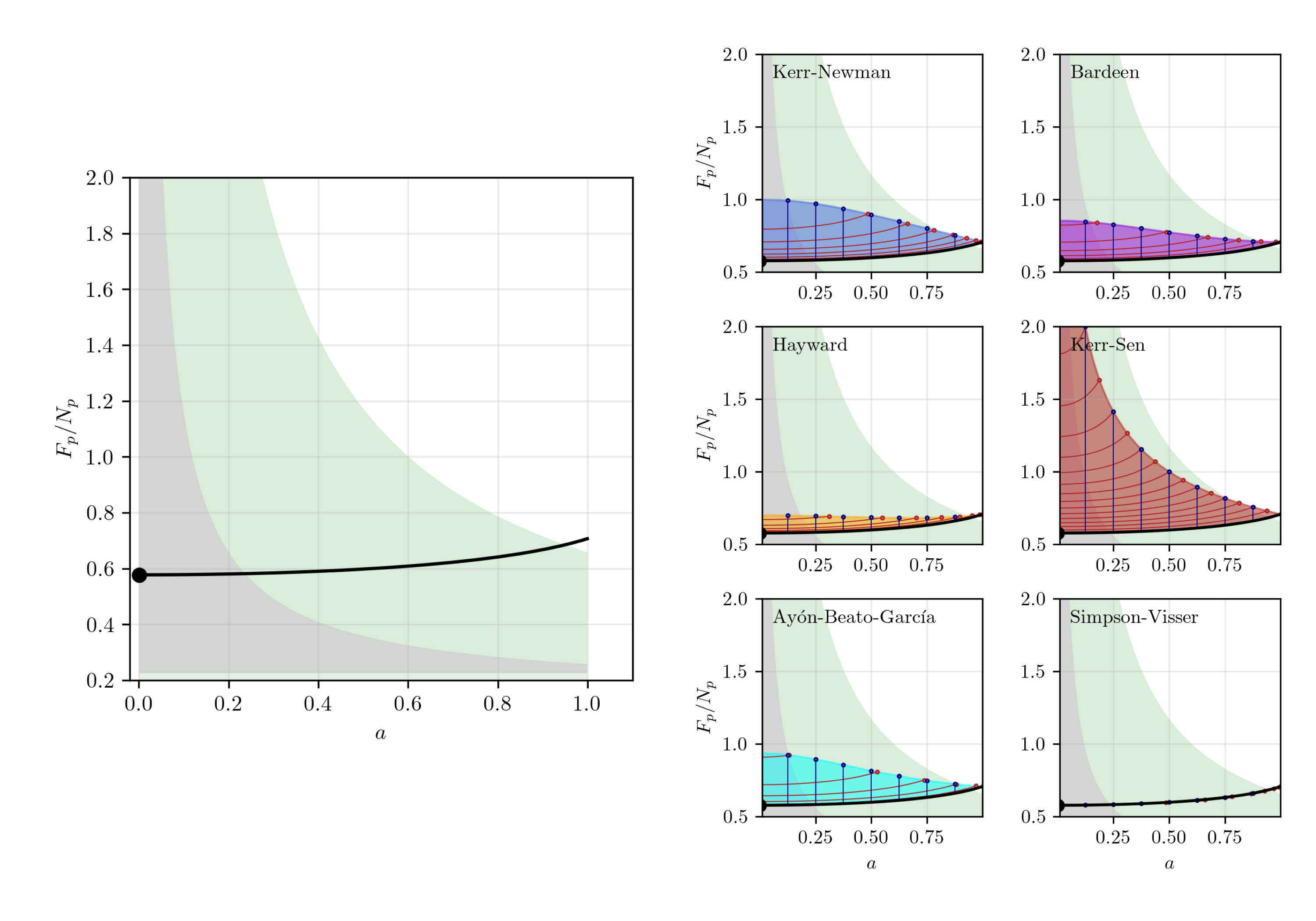}
    \caption{Shown here is a parameter space that captures frame-dragging in BH spacetimes, with the ratio of the metric functions $F/N$ evaluated at the complete photon sphere on the $y-$axis and the BH spin on the $x-$axis. In all panels, the green region represents the constraint imposed on this parameter space by the 2017 EHT shadow size measurement of M87$^*$. The gray band represents the region in which the the azimuthal shift lies within $10\%$ of its Schwarzchild value, $\delta_{\rm p} = \pi$, as a demonstrative example of a future measurement. The Kerr spacetime is represented by a solid black line in this parameter space whereas the Schwarzschild spacetime occupies a single point (black dot). Various alternative spinning spacetimes occupy different shaded regions. Blue lines represent constant-spin curves whereas the red lines represent constant-charge curves.}
    \label{fig:fig2_phi_measurement}
\end{figure*}

The polar value of the azimuthal shift, $\delta_0$, is an interesting observable since it dependeds on the ratio \( F/N \). This makes it a valuable tool to probe the effect of frame-dragging, using light. To compare, the other photon ring observables explored so far primarily offer insights into the function \( N \) and its derivative.

Equation \eqref{eq:Polar_Crit_Params} suggests that combining a future potential measurement of the azimuthal shift with the current M87$^*$ shadow size measurement provides a novel test of frame-dragging in BH spacetimes. More specifically, with these measurements, it is possible to impose joint agnostic constraints on the ratio \( F/N \) and the BH spin $a$, for this class of spacetime geometries, as described in Fig. \ref{fig:fig2_phi_measurement}. In each panel, the green shaded region corresponds to the region consistent with the M87$^*$ shadow size measurement. For demonstrative purposes, the gray region shows the part of this parameter space where the azimuthal shift lies within $10\%$ of the Schwarzschild value ($\delta = \pi$). Illustrated also are the maximal possible ranges of \( F/N \) as a function of spin for various alternative spinning BH spacetimes, with the Kerr spacetime represented by the solid black line and Schwarzschild spacetime by the black dot. This figure demonstrates how potential observations of the azimuthal shift provide inklings of frame-dragging effects caused by the compact object at the center of M87. 

Each subplot on the right side of this figure displays the valid predicted regions for each alternative spacetime. The current constraints, based on the EHT shadow size observation (indicated by the green-shaded regions), allow us to exclude only a small portion of the parameter space. Specifically, this applies to the nearly extremal spin regions close to \(a = 1\), which do not overlap with the green-shaded area.

As previously discussed, potential future measurements of the azimuthal shift could substantially narrow the observationally valid regions. This is represented by the grey band in the figure, which corresponds to the Schwarzschild spacetime value of the azimuthal shift with a \(10\%\) error margin. The overlap of this grey band with the shaded regions for each spacetime defines the further constrained valid regions, offering significantly tighter observational bounds compared to the current constraints. It is worth mentioning that we are only using the Schwarzschild value of azimuthal shift as a representative and we are not claiming that this is the true value for $M87*$.

Although the grey and green regions in these subplots suggest that the azimuthal shift measurement primarily constrains the spin parameter rather than \(F/N\), this outcome arises from the inherent nature of the azimuthal shift. As shown in \eqref{eq:Polar_Crit_Params}, when the spin parameter is zero, the azimuthal shift is exactly \(\pi\), rendering it uninformative about \(F/N\). However, for other values of the azimuthal shift, the situation differs. In such cases, the measurement provides constraints on \(F/N\) as a function of the spin parameter, offering valuable insights into the underlying spacetime structure.
 
A comprehensive treatment of the general expressions for observers at small inclinations is provided in the Appendix \ref{app: small inclination}. Here, we summarize the key signatures associated with such inclinations.

As the observer's inclination shifts from \(0^\circ\) to a small angle (e.g., \(17^\circ\) for the case of M87), the critical photon radius is no longer solely determined by Equation \eqref{eq:r polar}. Instead, a correction term, dependent on the inclination angle, is introduced. Consequently, the critical radius is influenced not only by the function \(N\) and its first derivative but also by the function \(F\) and its derivative. However, this deviation from the polar value remains minor, with a relative correction of at most \(\delta r / r_0 \approx 10^{-4}\) in the case of a maximally spinning Kerr spacetime. Accordingly, all results are computed up to the first order of \(\delta r / r_0\).

Consequently, considering the first observable, the shadow size deviates from its polar value. For M87 with an inclination of \(17^\circ\), this deviation is at most \(\delta \rho / \rho_0 \approx 10^{-3}\) for a maximally spinning Kerr black hole. Detecting this deviation could provide insights into the values of the function \(F\) and its derivative, as well as the second derivative of \(N\).

Other observables such as the critical parameters are also influenced by the inclination angle, except for the time delay, which remains consistent with its polar value to first-order corrections. As a result, the time delay can be directly inferred from the shadow size and spin parameter and remains indirectly related to the inclination angle and the metric functions \(F\) and \(N\).
Therefore, combining this measurement with the shadow size provides a direct constraint on the spin parameter.

Finally, it is worth emphasizing that by leveraging these combined observational constraints, we can construct a more comprehensive understanding of the spacetime geometry. This includes uncovering intricate details of the ergosphere, event horizon, spin parameter, and the behavior of spacetime in the vicinity of compact astrophysical objects such as M87$^*$.

We will consider the variation of the critical parameters obtained here for a large obsever inclination angle, as relevant for images of Sgr A*, in our companion paper, Ref. \cite{Kumar+2024}.

\section{Conclusion}
In this work, we employ a general, non-perturbative, and non-parametric axisymmetric framework to investigate potential future tests of gravity. While this spacetime model may not capture the most general integrable axisymmetric solution, it represents a broad class of theoretically significant alternative spacetimes. This formalism, introduced in \citet{Salehi+2023}, reformulates the general axisymmetric spacetime originally proposed in \citet{Johannsen:2013szh}. This reformulation is particularly valuable, as noted in \cite{Salehi+2023}, because the framework’s free functions carry important physical interpretations. For example, the function \( N \) relates to the existence and shape of the event horizon, while the function \( F \) is associated with the ergosphere. Consequently, this general spacetime model enables the calculation and interpretation of various gravity tests with greater clarity.

Additionally, we examine observables such as the shape and size of the BH shadow, as well as the characteristics of higher-order images, including the Lyapunov exponent, time delay, and azimuthal shift between successive rings. Previous calculations of these observables were either specific to Kerr spacetime, as in \cite{Gralla:2019drh}, or applied to a general framework but limited to polar observers and fewer observables, as in \cite{Salehi+2023}. By extending the framework to a generalized spacetime and accommodating arbitrarily inclined observers, our approach offers a comprehensive method for examining spacetime structure, spin parameters, and observer inclination, providing a broader and more versatile foundation for testing the nature of gravity.

We begin by deriving expressions for the shadow size, Lyapunov exponent, time delay, and azimuthal shift for a generally inclined observer. Following this, we focus on the polar values of time delay and azimuthal shift, which are particularly relevant for observations of M87$^*$. Although current observational limitations prevent the precise measurement of these quantities, we anticipate that advancements in resolution and sensitivity will make such observations feasible in the future. Additionally, we reserve a more detailed investigation into the interpretations of these measurements at higher inclinations for future work.

We apply this framework to analyze the Event Horizon Telescope (EHT) observations of M87$^*$'s shadow size. Previous work by \cite{Salehi+2023} explored constraints on the Lyapunov exponent for a polar observer within this general framework. Here, we expand the analysis to include two additional critical parameters, demonstrating that the time delay observed by polar observers serves as a degenerate measure of both spin and shadow size. This result indicates that time delay is not directly tied to the underlying spacetime geometry but rather depends implicitly on it through the shadow size. By contrast, the azimuthal shift observed by a polar observer is directly influenced by spacetime geometry, specifically by the ratio \( F/N \) at the critical photon ring, the spin parameter, and the shadow size.

As a result, future observations that combine precise measurements of shadow size and time delay can serve as direct constraints on the spin value. Additionally, simultaneous measurements of shadow size, time delay, and azimuthal shift could provide direct insight into spacetime geometry, particularly regarding the value of \( F/N \) at critical radii. Notably, these potential observations suggest both qualitative and quantitative insights; for instance, the presence of a non-zero value of \( F_p/N_p \) is intrinsically linked to the existence of frame-dragging and the ergoregion.

Although we introduced a general framework, it is restricted to a broad but not fully general class of axisymmetric spacetimes. This excludes non-integrable spacetimes, examples of which are present in the literature. Additionally, we do not account for non-stationary spacetimes, though this limitation is unlikely to significantly affect imaging experiments such as those conducted by the EHT and future instruments.

Another key limitation of our work is the focus on polar observer's viewpoint in our physical interpretations. While the calculations were performed for a general observer, we primarily explored scenarios relevant to polar observers. This provides an appropriate initial approximation for M87$^*$, but it may not be directly applicable to Sgr A$^*$, where the inclination remains uncertain.

Additionally, throughout the entire study, we have deliberately avoided exploring the astrophysical complexities that might arise in observing and measuring these quantities. The observables we examine here are purely gravitational effects related to very high-order photon rings, which are currently beyond our observational reach. However, measuring low-order rings may become feasible in the near future. Therefore, it is important to consider the astrophysical effects that measuring these quantities for low-order photon rings might introduce, which is a context not addressed in this work.

\begin{acknowledgments}
The Perimeter Institute for Theoretical Physics partially supported this work. Funding for research at the institute is provided by the Department of Innovation, Science and Economic Development Canada, and the Ministry of Economic Development, Job Creation and Trade of Ontario, both of which are branches of the Government of Canada. 
A.E.B. expresses gratitude to the Delaney Family for their generous financial backing through the Delaney Family John A. Wheeler Chair at Perimeter Institute. Additionally, A.E.B. receives further financial support for this research through a Discovery Grant from the Natural Sciences and Engineering Research Council of Canada.
PK acknowledges support from grants from the Gordon and Betty Moore Foundation (GBMF-8273) and the John Templeton Foundation (\#62286) to the Black Hole Initiative at Harvard University. RKW's research is supported by the Fulbright-Nehru Postdoctoral Research
Fellowship (Award No. 2847/FNPDR/2022) from the United States-India Educational Foundation.
\end{acknowledgments}

 \clearpage
 \newpage
\begin{appendix}
\section{Measuring Azimuthal Shift in Kerr Spacetime}
\label{app:Kerr_phi}
Assuming the polar observer the azimuthal shift is governed by \eqref{eq:Polar_Crit_Params}, which is stated in a non-purturbative and non-parametric framework. However, assuming a specific spacetime for instance Kerr spacetime, gives (see Table 1 in \cite{Salehi+2023}),

\begin{align} 
\label{eq:Kerr}
N = &\ \frac{r \sqrt{\Delta}}{r^2+a^2}\\
F  = &\  \frac{r}{r^2+a^2},
\end{align} 
where $\Delta = r^2 - 2 r +a^2$. Additionally, the critical radius viewed by a polar observer in Kerr spacetime is equal to (see equation (C9)in \cite{Salehi+2023}:
\begin{equation}
\begin{split}
     r & = M + 2M \sqrt{1-\frac{a^2}{3M^2}}\\ & \times
    \cos\left[
    \frac{1}{3} \cos^{-1}\left(
    \frac{1-a^2/M^2}{(1-a^2/3M)^{3/2}}
    \right)
    \right].
    \label{eq:Kerr_rp}
\end{split}  
\end{equation}
Therefore the azimuthal shift in this spacetime is equal to:

\begin{align} \label{eq:varphi_p_Def_polar_Kerr}
\varphi_{\mathrm{p}} =\ \frac{4  a}{\sqrt{ \left(\frac{1}{\partial_r N}\right)^2-a^2}}  \left[\frac{1}{\sqrt{r^2- 2r+a^2}} \frac{1}{\partial_r N} - 1 \right]K \left( a^2 \partial_r N^2 \right),
\end{align} 
which, $\partial_r N$ is the first derivative of \eqref{eq:Kerr} at the critical radius. Thus, the azimuthal shift directly gives information on the value of the spin parameter. Similar arguments can be made assuming other spacetimes as well.

\section{Polar Azimuthal Shift Plot}
\label{app:f/N plot}
In this appendix we explicitly explain the details of Fig.2. First we start with the green shaded region demonstrating the EHT observations of Shadow size measuremnt for M87$^*$. In order to do so, we start by rearranging equation \eqref{eq:Polar_Crit_Params} to isolate $F_P/N_P$,
\begin{align} \label{eq:F/N and spin}
\frac{F}{N}  = \frac{1}{\mathscr{I}}\left[\frac{\varphi_{\mathrm{p}} - \pi}{a G^1_{\vartheta}(m) } + 1 \right].
\end{align}
Now $F_P/N_P$ is a function of azimuthal shift, spin, and shadow size. By incorporating the EHT measurement of shadow size, this dependence reduces to only the azimuthal shift, \(\phi\), and spin, \(a\). To estimate \( \frac{F_P}{N_P} \) solely as a function of spin, a measurement of the azimuthal shift is required, which is not available at the moment. However, by using the Kerr values of \(\phi\), we can plot the green regions in Fig.2. These green regions overlap almost completely, making them difficult to distinguish.

A similar approach was applied to the gray measurements. Instead of using the full range of azimuthal shift values in Kerr spacetime, the Schwarzschild value of \(\pi\) with a \(10\%\) margin of error was used as an illustrative example of a potential future measurement of \(\phi\). We emphasize that this does not imply a necessary proximity to Schwarzschild spacetime; rather, it serves as a representative example.

\section{Small inclinations} 
\label{app: small inclination}

Here, we examine the observational signature of introducing a small inclination angle, which is particularly relevant to M87$^*$ given its estimated inclination of $17^\circ$ (\cite{EHTC+2019f}). Specifically, we investigate how a slight inclination affects observables such as the shadow size and various critical parameters. The unknown inclination angle for Sgr A$^*$ introduce complexities that make its analysis beyond the scope of this study, which we leave this for future work. 

We analyzed the case for observers present at precisely zero inclination in Sec. \ref{Sec-3} above. For such polar observers, the shadow boundary curve and the critical parameters do not vary over the image plane. This is because the photon shell reduces to a single complete photon sphere located at $r=r_0$.

For observers with small inclination angles, \( \mathscr{i} \ll \pi \), a band of photon shell radii contribute to the formation of the critical curve. These radii shift slightly from its polar value (see eq. 4 of Ref. \cite{Salehi+2023}) by an amount we denote as \( \Delta r \). This shift can be calculated by solving the angular potential equation for small inclinations, up to first order
\begin{equation}
\Delta r = \left[\frac{\partial_r F N-\partial_r N F }{N \partial^2 _r N }\right]_0\mathscr{i}.
\end{equation}
As in Sec. \ref{Sec-3}, the subscript ``0'' indicates that the functions are evaluated at the radius of the complete photon sphere, $r=r_0$.

Consequently, the critical values of the conserved quantities for these null orbits also differ from their polar values, to first order, by,
\begin{equation}
\begin{split}
\Delta \xi & =  - \frac{1}{a}\left[\frac{r \partial^2_r N  }{N \partial_r F-F \partial_r N }\right]_0\Delta r, \;\;\;  \\
\Delta \mathscr{I} & =  \left[\frac{1}{\partial_r N}\frac{F  \partial^2_r N  }{N\partial_r F-F \partial_r N}\right]_0\Delta r,
\end{split}
\label{eq:delta xi and delta eta}
\end{equation}

Therefore the shadow size, using Eq. \eqref{eq:rho and psi}, also changes to
\begin{eqnarray}
\label{eq:shadow size}
\rho^2 = \frac{1}{\left(\partial_r N_0\right)^2 } + 2 \frac{\Delta\mathscr{I}}{\partial_r N_0}  + 2  a(\Delta \xi)
\end{eqnarray}

This demonstrates that the shadow size shifts slightly from its polar value by an amount proportional to \( \Delta r \), and ultimately depends on the inclination angle and the geometry of spacetime. Furthermore, this suggests that the shadow size measurement, which was previously informative only about the first derivative of the function \( N(r, a) \) at the critical radius in the polar observer scenario (see Figure 1 and equation (18) in \cite{Salehi+2023}), can now also be sensitive to the function \( F \) and its derivatives. As a result, it can provide insights into the characteristics and nature of the ergosphere.

We now proceed to calculate the critical exponents. We can obtain the solutions to the polar turning point, which is the only remaining input necessary for this calculation, for a nearly-polar observer as being (Eq. \ref{eq:u_sol})
\begin{equation}
\begin{split}
 u_+ = 1, \;\;\; 
 u_- & = -\frac{1}{a^2} \left( \eta_0+ \Delta \eta  \right)\,,
\end{split}
\label{eq:u plus and minus nearly polar}
\end{equation}
where $\eta_0$ is the Carter constant for the zero angular momentum SNG \eqref{eq:Polar_Turning_Points_Polar_Obs} and $\delta \eta$ is obtained from eq. \ref{eq:Carter_Consts} as being $\Delta \eta = 2 \mathscr{I}_0 \Delta \mathscr{I} + 2 a \Delta \xi$.

Utilizing these expressions for $u_\pm$ in Eqs. \eqref{eq:MinoTime} and \eqref{eq:hat_G_t} results in,
\begin{equation}
\begin{split}
 \hat{G}_\vartheta & = \hat{G}_{\vartheta; 0} - \frac{E(m)}{\sqrt{\eta_0} }   \frac{\Delta \eta}{\eta_0 +a^2} , \\
 \hat{G}_t & = \hat{G}_{t;0} + \frac{1}{a^2}\left( \frac{\eta_0}{\mathscr{I}_0^2}E(m)-  K(m) \right) \frac{\Delta \eta }{ \sqrt{\eta_0}},
\end{split}
\label{eq: G nearly polar}
\end{equation}
where $m = u_{-;0} = - a^2/\eta_0$ and $\hat{G}_{\vartheta; 0}$ and $\hat{G}_{t; 0}$ are defined in Eq. \eqref{eq:polar G_theta and G_t}. However, the situation differs for $ G_\varphi$. For nearly polar observers, $\hat{G}_\varphi$ diverges and becomes undefined. Despite this, the product $\xi_p \hat{G}_\varphi$ remains finite and is equal to $ 2 \pi$, as discussed in Eq. 66 of Ref. \cite{Gralla:2019drh}. 

The demagnification exponent exponent can now derived from Eq. \eqref{eq:gamma_p_Def} by expanding it for nearly polar observers, to yield
\begin{equation}
    \Delta \gamma = (\Delta \kappa) \hat{G}_{\vartheta;0} + \kappa_{0}(\Delta\hat{G}_{\vartheta})\,,
\end{equation}
where $\Delta \hat{G}_\vartheta := \hat{G}_\vartheta - \hat{G}_{\vartheta; 0}$ can be read off from Eq. \eqref{eq: G nearly polar} above. 

Since the phase space Mino time Lyapunov exponent, $\kappa$, is given as,
\begin{equation}
    \kappa =\frac{r N}{B}  \sqrt{\mathscr{I}\partial_r^2\left( \frac{r-aF \xi}{N}\right)}\,,
\end{equation}
its value at the complete photon sphere, $\kappa_0$, is seen to be
\begin{equation}
    \kappa_0 = \frac{N^{3/2}}{\left( \partial_r N\right)^2}  \sqrt{\frac{-\partial_r^2 N}{B^2}}\,,
\end{equation}
and its deviation due to small inclination, $\Delta\kappa$, is,
\begin{equation}
   \frac{\Delta \kappa}{\kappa_0} = \left[\frac{1}2{}\frac{\partial_r^2 F}{\partial_r^2 N}\right]_0 a\Delta \xi + \Delta \mathscr{I} + \left[\frac{\partial_r \ln \left( N^3/(B^2 \partial_r^2 N)\right)}{2 B}\right]_0 \Delta r.
\end{equation}
This is the only observable in our work which is dependent on the function $B$ and its derivative as well as the third derivative of function $N$.

In order to derive the devation of the time delay from its polar value  similarly one can use Eq. \eqref{eq:tp_Def} and expand it to first order, which yields to,
\begin{equation}
    \Delta \tau = (\Delta J) \hat{G}_\vartheta + J_0 (\Delta \hat{G}_\vartheta) + a^2 (\Delta \hat{G}_t),
\end{equation}
where \( J_0 \) is the polar value of the function \( J(r) = \left[\frac{r}{N} \mathscr{I} + a \xi - a^2\right] \), and is explicitly given as
\begin{equation}
    J_0 = \mathscr{I}_0^2 - a^2.
\end{equation}
Additionally, 
\begin{equation}
    \Delta J = \left[\mathscr{I}_0(\Delta I) + a(\Delta \xi)\right] = \frac{\Delta \eta}{2}.
\end{equation}

Substituting these expressions and using \eqref{eq: G nearly polar}, we find
\begin{equation}
    \Delta \tau = 0.
\end{equation}

Lastly, the azimuthal shift can be obtained using \eqref{eq:varphi_p_Def} and expanding it up to first order which gives,
\begin{equation}
    \Delta \delta =  \mathcal{F}(\Delta \hat{G}_\vartheta)  +  (\Delta \mathcal{F})\ \hat{G}_{\vartheta; 0}
\end{equation}
where $\mathcal{F}$ and $\Delta \mathcal{F}$ are govern by the following equations,
\begin{align}
    \mathcal{F} & = a \left[\frac{F}{N \partial N} - 1\right],\\
    \Delta \mathcal{F} & = a \left[ 
    \frac{\partial_r F N-\partial_r N F}{N^2 \partial_r N} \right]_0\Delta r +
    a\left[\frac{F}{N}\right]_0 \Delta \mathscr{I}.
\end{align}

It is worth highlighting that the
$\xi_{\mathrm{p}} \hat{G}_\varphi$ term goes to $ \pi$ when $\mathscr{i} \rightarrow 0$.

Therefore, it does not contribute to $\Delta \delta$. Consequently, the final expression reduces to,
\begin{equation}
    \Delta \delta = \frac{2a^2}{N_0^2 \sqrt{\eta_0}} \left( \pm K(m) \mathscr{I}_0 (\Delta r) - E(m) (F_0-N_0 \partial_r N_0) N_0 \frac{\Delta \mathscr{I}}{\mathscr{I}_0}\right)\,,
\end{equation}
where the $\pm$ sign in the first term depends on the sign of the expression $N \partial_r F - F \partial_r N$.

\end{appendix}

\bibliography{PRD.bib}


\end{document}